\documentstyle[twocolumn,prb,aps,epsf,psfig]{revtex}

\begin{document}

\input{psfig}



\twocolumn[\hsize\textwidth\columnwidth\hsize\csname @twocolumnfalse\endcsname 
\title{\bf Non-uniqueness of late-time scaling states in spinodal
decomposition} 
\author{A.J. Wagner and C.E. Scott}

\address{ Department of Material Science and Engineering,\\
Massachusetts Institute of Technology,\\ 77 Massachusetts Avenue,
Cambridge, MA 02139, U.S.A.  } \date{\today} \maketitle

\begin{abstract}
\begin{center} \bf Abstract \end{center}
In this letter we show that the late-time scaling state in spinodal
decomposition is not unique. We performed lattice Boltzmann
simulations of the phase-ordering of a 50\%-50\% binary mixture using
as initial conditions for the phase-ordering both a symmetric
morphology that was created by symmetric spinodal decomposition and a
morphology of one phase dispersed in the other, created by viscoelastic
spinodal decomposition. We found two different growth laws at late
times, although both simulations differ only in the early time
dynamics. The new scaling state consists of dispersed droplets. The
growth law associated with this scaling state is consistent with a
$L\sim t^{1/2}$ scaling law.
\end{abstract}
\vspace{0.3cm}

]


\section{Introduction}
The phase-ordering in binary fluids following spinodal decomposition
has been the focus of considerable interest in the recent
past\cite{Bray,prl,Kendon2,Kendon1}. A number of different scaling
regimes with different growth laws have been identified with the help
of scaling arguments, but little emphasis has been placed on
morphological effects\cite{prl,mecke}. Unusual asymmetric morphologies
are often seen in the early stages of viscoelastic spinodal
decomposition \cite{Bhattacharya,Haas,Onuki}, but it has been argued
that viscoelastic effects are unimportant at late times and therefore
only one true late-time phase-ordering scaling state
exists\cite{Bhattacharya}. The term ``late-time scaling state'' is not
clearly defined since a number of scaling states that cross over at
certain length-scales have been predicted and observed. In the present
context, we take a ``late-stage'' scaling state to mean a fully
phase-separated state in which the time dependence of a typical length
scale $L(t)$ has converged to a growth law of the form $L(t)\sim
t^\alpha$. Even though the existence of these scaling states in a
strict sense has been questioned\cite{prl,grant}, the concept remains
useful to understanding phase-ordering dynamics in binary fluids.

The argument of the unimportance of viscoelasticity in the late-stage
growth, as well as many of the scaling arguments, rely on an assumption
that is often implicitly made: for the phase-ordering of symmetric
binary fluid mixtures only one universal late-time scaling state
exists\cite{Bray}. In this letter, however, we show that this
assumption is not valid, and that therefore the early time dynamics are
very important in selecting the late-time scaling state. The
possibility of the non-uniqueness of the scaling state was first
suggested by A. Rutenberg, although at the time no persuasive
numerical evidence could be found\cite{Rutenberg}.

\section{Numerical method}
For the simulations we use the viscoelastic two-component
lattice Boltzmann simulation introduced in an earlier
paper\cite{vedrop}. Briefly, in lattice Boltzmann simulations
densities $f_i$ that are associated with velocities $v_i$ are streamed
on a lattice according to the lattice Boltzmann equation
\begin{eqnarray}
&&f_i({\bf x}+{\bf v}_i \Delta t,t+\Delta t)=\nonumber\\
&&f({\bf x},t)
+ \Delta t \sum_j \Lambda_{ij} \left[f_j^0({\bf x},t)-f_j({\bf x},t)\right]
\end{eqnarray}
where $f_i^0$ is an equilibrium distribution, $\Lambda_{ij}$ is a
collision matrix, and ${\bf v}_i \Delta t$ is a lattice vector. The
velocity set for our simulation consists of 17 velocities given by
$\{(0,0)$, $(1,0)$, $(0,1)$, $(-1,0)$, $(0,-1)$, $(1,1)$, $(-1,1)$,
$(-1,-1)$, $(1,-1)$, $(1,0)$, $(0,1)$, $(-1,0)$, $(0,-1)$, $(1,1)$,
$(-1,1)$, $(-1,-1)$, $(1,-1)\}$. Note that the last 8 velocities are
the same as the previous eight velocities. This duplicity allows
the simulation to have two independent stresses which represent a
viscoelastic and a purely viscous contribution to the total stress
tensor. The two contributions are used to produce a Jeffrey's model
for the stress (see eqn. (\ref{jeffrey})). The algorithm is required
to conserve mass and momentum, but not energy. Energy conservation is
replaced by a condition of constant temperature. The macroscopic
density $\rho$ and velocity ${\bf u}$ are defined as
\begin{equation}
\rho=\sum_i f_i \;\;\; \rho {\bf u}=\sum_i f_i {\bf v}_i.
\end{equation}
To simulate a two-component mixture we introduce a second lattice
Boltzmann equation for densities $g_i$
\begin{eqnarray}
&&g_i({\bf x}+{\bf v}_i \Delta t,t+\Delta t)=\nonumber\\
&&g({\bf x},t)
+ \frac{\Delta t}{\tau} \left[g_i^0({\bf x},t)-g_i({\bf x},t)\right]
\end{eqnarray}
where we choose a diagonal collision matrix with a single relaxation
time $\tau$. These densities are only defined on the first nine
velocities ${\bf v}_i$. The density difference $\phi=\rho_1-\rho_2$ of
the two components is given by
\begin{equation}
\phi = \sum_i g_i
\end{equation} 
and the total density is given by $\rho=\rho_1+\rho_2$. By choosing
appropriate equilibrium distributions and an appropriate collision
matrix, we ensure that the following partial differential equations are
being simulated up to second order in the derivatives but assuming
that the relaxation of the viscoelastic stress $\sigma$ is slow ($\theta \sim
1/\sqrt{\epsilon}$):
\begin{eqnarray}
\partial_t \rho + \partial_{\bf x} (\rho {\bf u}) &=& 0\\
\rho \partial_t {\bf u} + \rho {\bf u}.\nabla {\bf u} &=&
-\partial_{\bf x} P + \partial_{\bf x} (\sigma_v+\sigma)\\
\sigma_v &=& \nu_\infty ( \nabla (\rho {\bf u}) + (\nabla
(\rho {\bf u}))^T - \nabla.{\bf u} \delta)\nonumber\\
&&+ \xi_\infty \nabla.{\bf u}\delta\\
\sigma + \theta(\phi) \sigma_{(1)} &=& -
(\nu_0(\phi)-\nu_\infty)(\nabla (\rho {\bf u})+ (\nabla (\rho {\bf u}))^T)\label{jeffrey}\\
\partial_t \phi + \partial_{\bf x}(\phi {\bf
u})&=& D \nabla^2 \mu + \nabla.((\phi/\rho)\nabla.(P-\sigma)) \label{diffusion}
\end{eqnarray}
where $\delta$ is the identity matrix, $\sigma$ is the viscoelastic
stress tensor, $\sigma_{(1)}=\partial_t \sigma + {\bf u}.\nabla
\sigma- \sigma.(\nabla {\bf u})-(\nabla {\bf u})^T \sigma$ is its
upper convected derivative, $P=0.5 \rho+0.007 (\nabla \phi \nabla \phi
- 0.5 \nabla \phi.\nabla \phi \delta- \phi \nabla^2 \phi \delta$ is
the pressure tensor, and $\mu=-0.55 \phi/\rho + 0.25 \ln
((\rho+\phi)/(\rho-\phi)) - 0,007 \nabla^2 \phi$ is the chemical
potential. The parameters $\nu_\infty$, $\xi_\infty$, and $\theta$ are
determined by the eigenvalues of the collision matrix. The values for
the parameters were $\Delta t=1$, $\tau=1$, $D=0.5$, $\xi_\infty=0.31$, and
$\nu_\infty=0.01$. For the low-viscosity phase we used $\theta=0.055$,
$\nu_0=0.013$ and for the viscoelastic phase $\theta=39.5$ and
$\nu_0=1.97$. For the symmetric simulations of Figure \ref{fig:comp}
we used $\theta=0.055$ and $\nu_0\approx \nu_\infty=0.075$.

\section{Simulations}
We performed simulations of critical spinodal decomposition of a
viscoelastic binary mixture in two dimensions where one component is
much more viscoelastic than the other. These simulations lead to the
usual morphologies in which the viscoelastic phase is connected and
the less viscoelastic phase is dispersed\cite{ve-decomp}. We performed
our simulations on a $256^2$ lattice and after about 1000 iterations
the less viscoelastic phase is completely dispersed, although the
domains are still highly deformed. We used this morphology as an
initial condition for a simulation where we make both components
purely viscous to examine the effects of initial conditions that are
not symmetric on symmetric binary fluid mixtures. This allows us to
distinguish the effect that the morphology created by
viscoelastic phase separation has from the effect of viscoelasticity
itself in the late-time phase-ordering process.

\begin{figure}
\begin{center}
\begin{minipage}{7cm}
\begin{minipage}{3cm}
\centerline{\psfig{figure=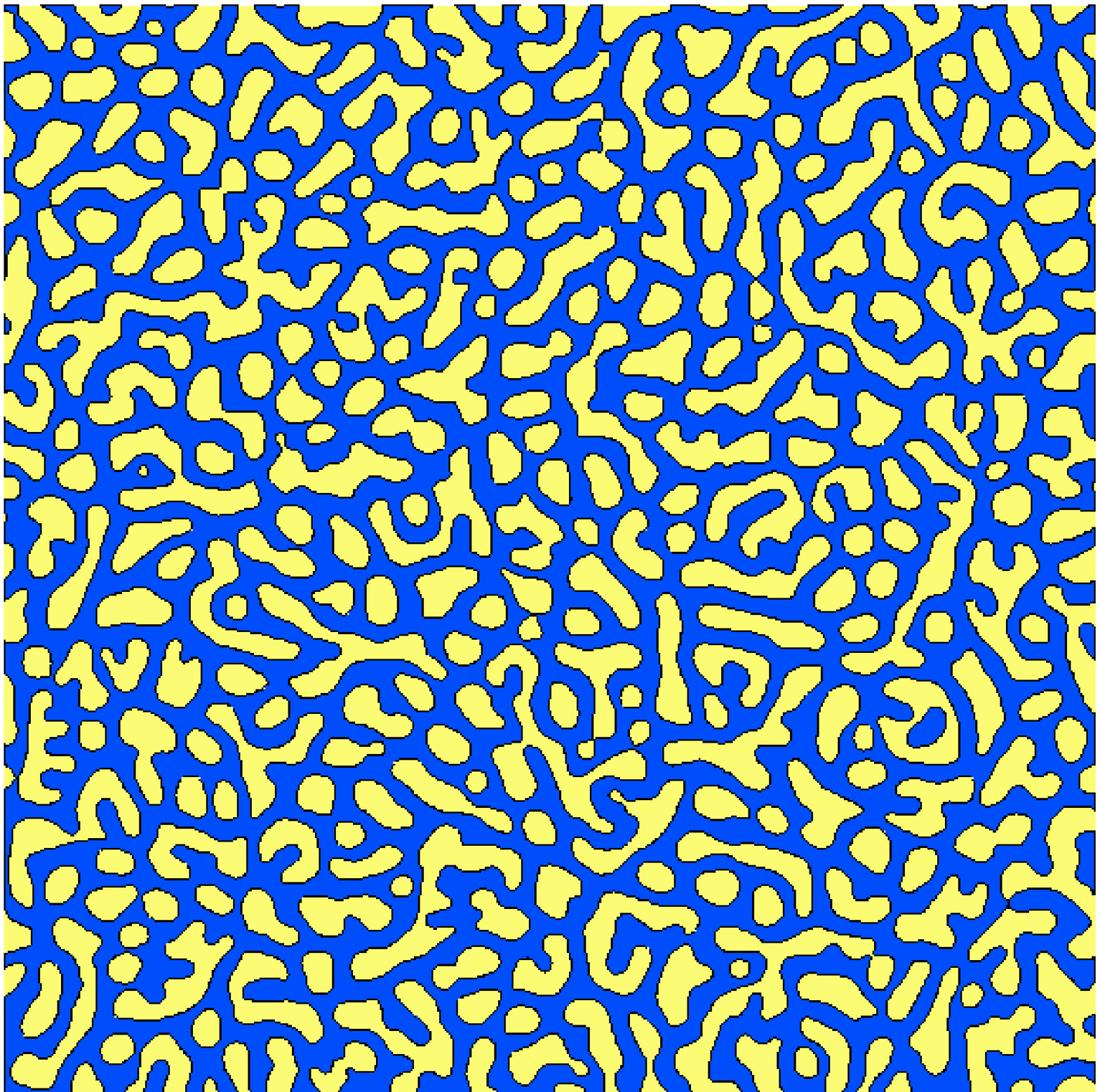,width=3cm}}
\end{minipage}
\hspace{0.3cm}
\begin{minipage}{3cm}
\centerline{\psfig{figure=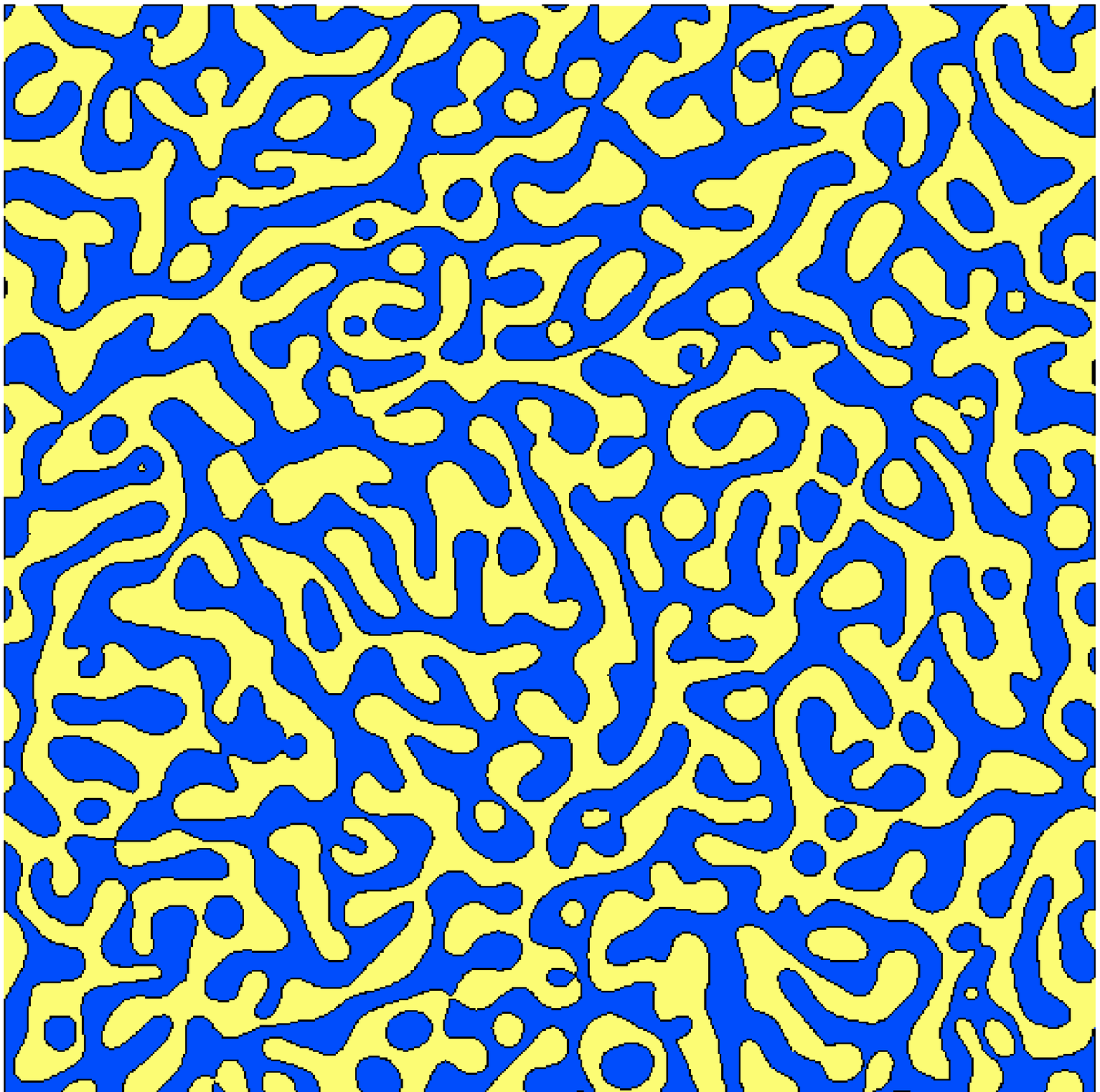,width=3cm}}
\end{minipage}\\
\begin{minipage}{3cm}
\centerline{\psfig{figure=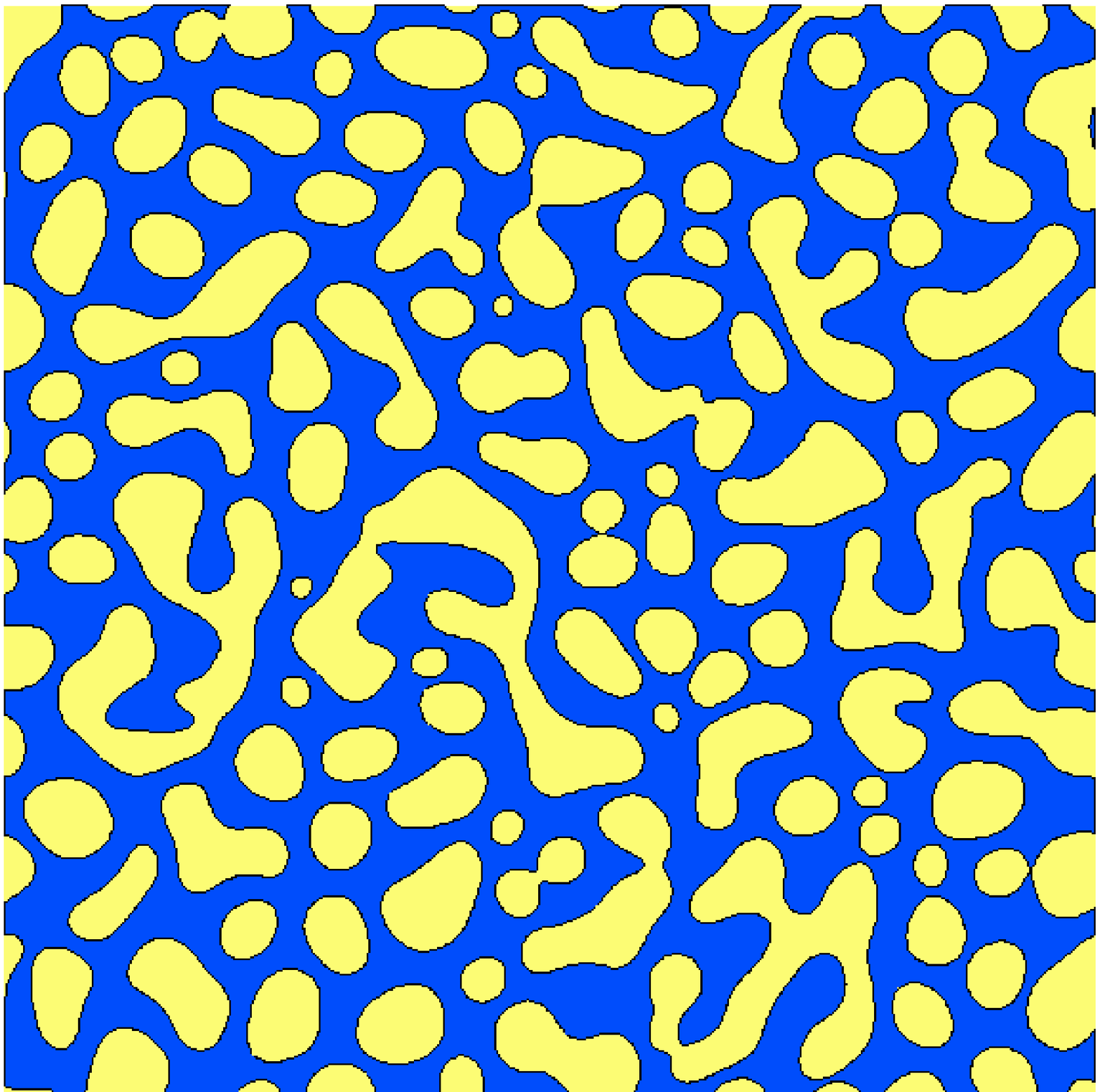,width=3cm}}
\end{minipage}
\hspace{0.3cm}
\begin{minipage}{3cm}
\centerline{\psfig{figure=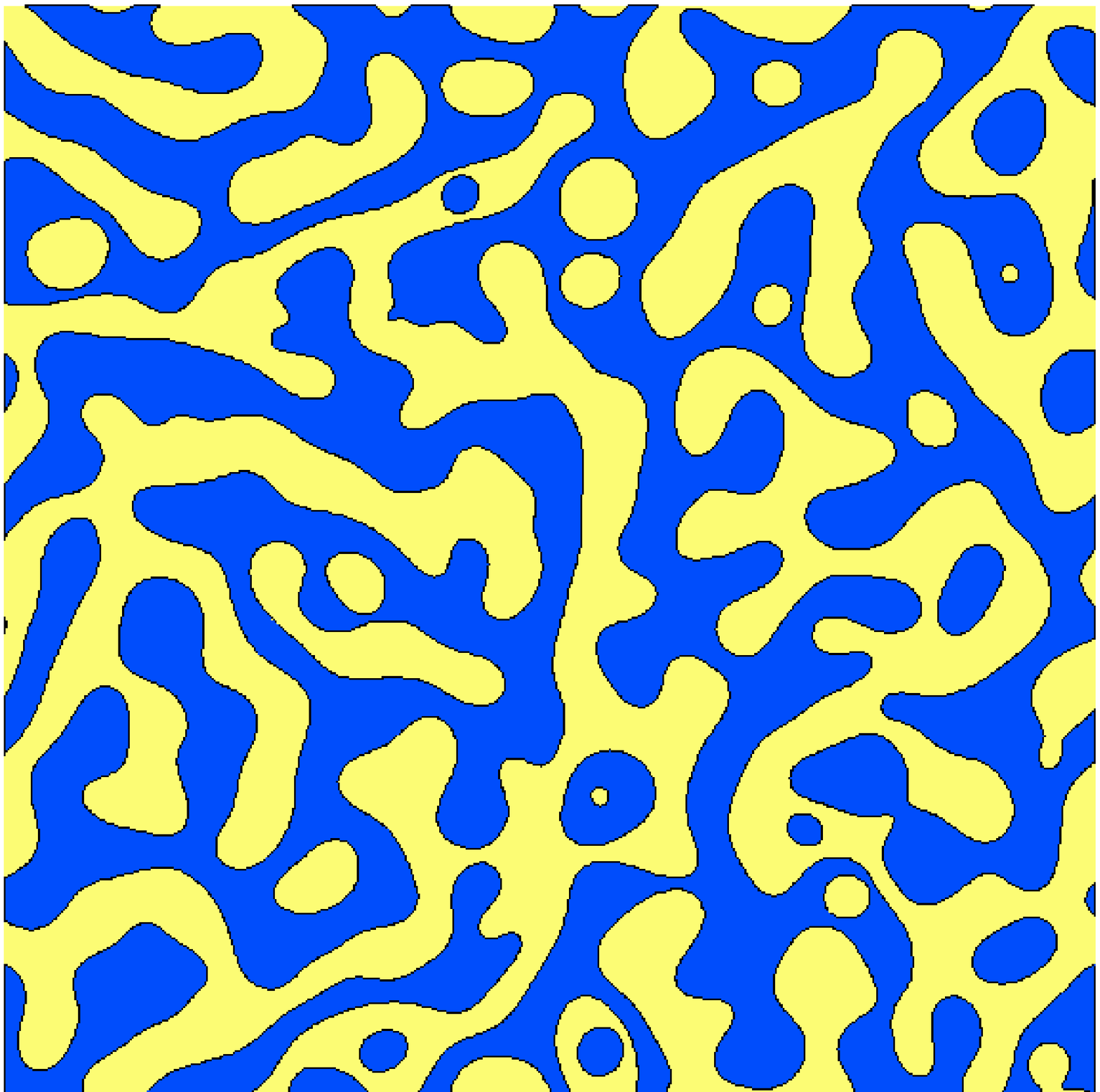,width=3cm}}
\end{minipage}\\
\begin{minipage}{3cm}
\centerline{\psfig{figure=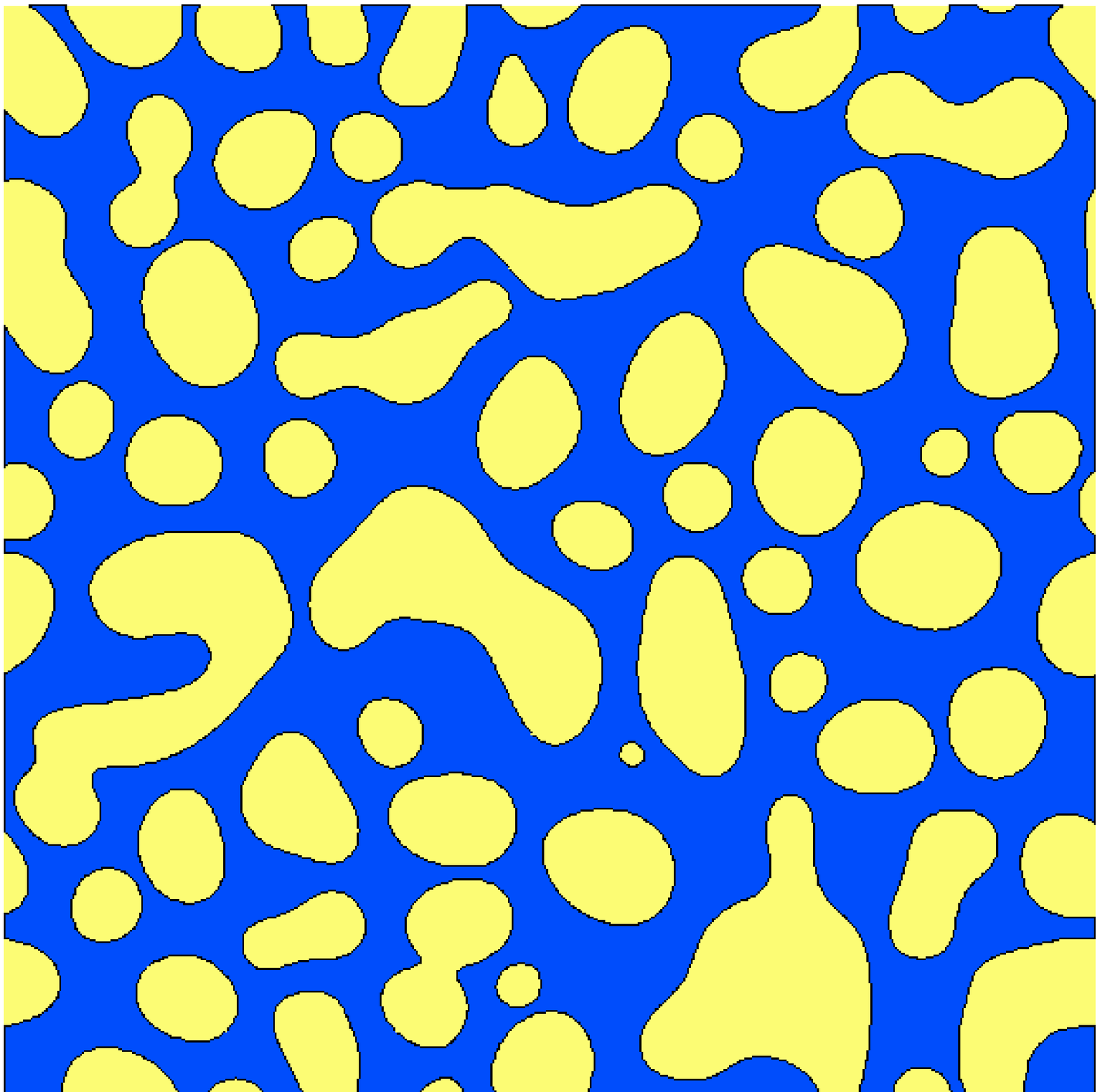,width=3cm}}
\end{minipage}
\hspace{0.3cm}
\begin{minipage}{3cm}
\centerline{\psfig{figure=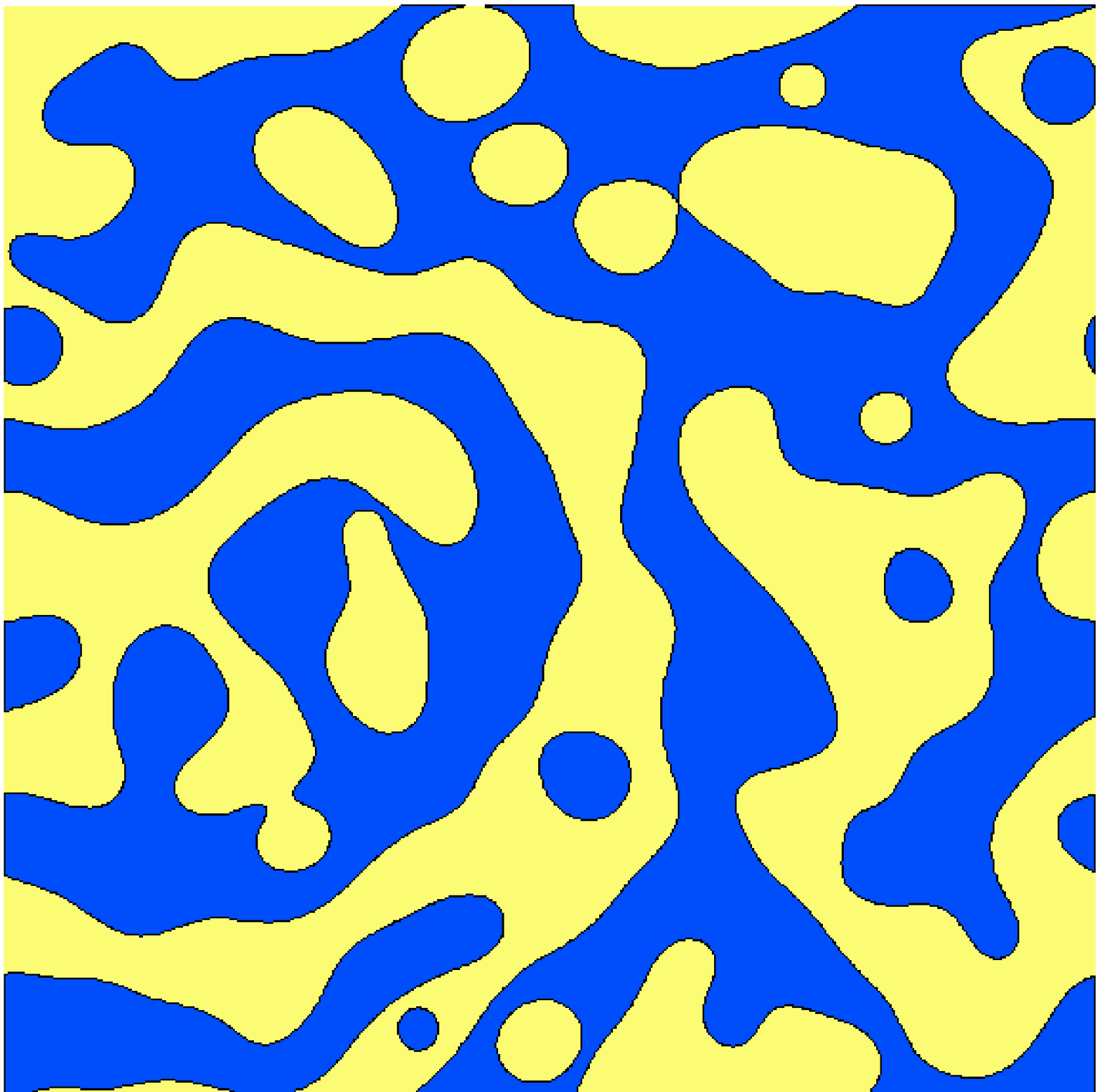,width=3cm}}
\end{minipage}\\
\begin{minipage}{3cm}
\centerline{\psfig{figure=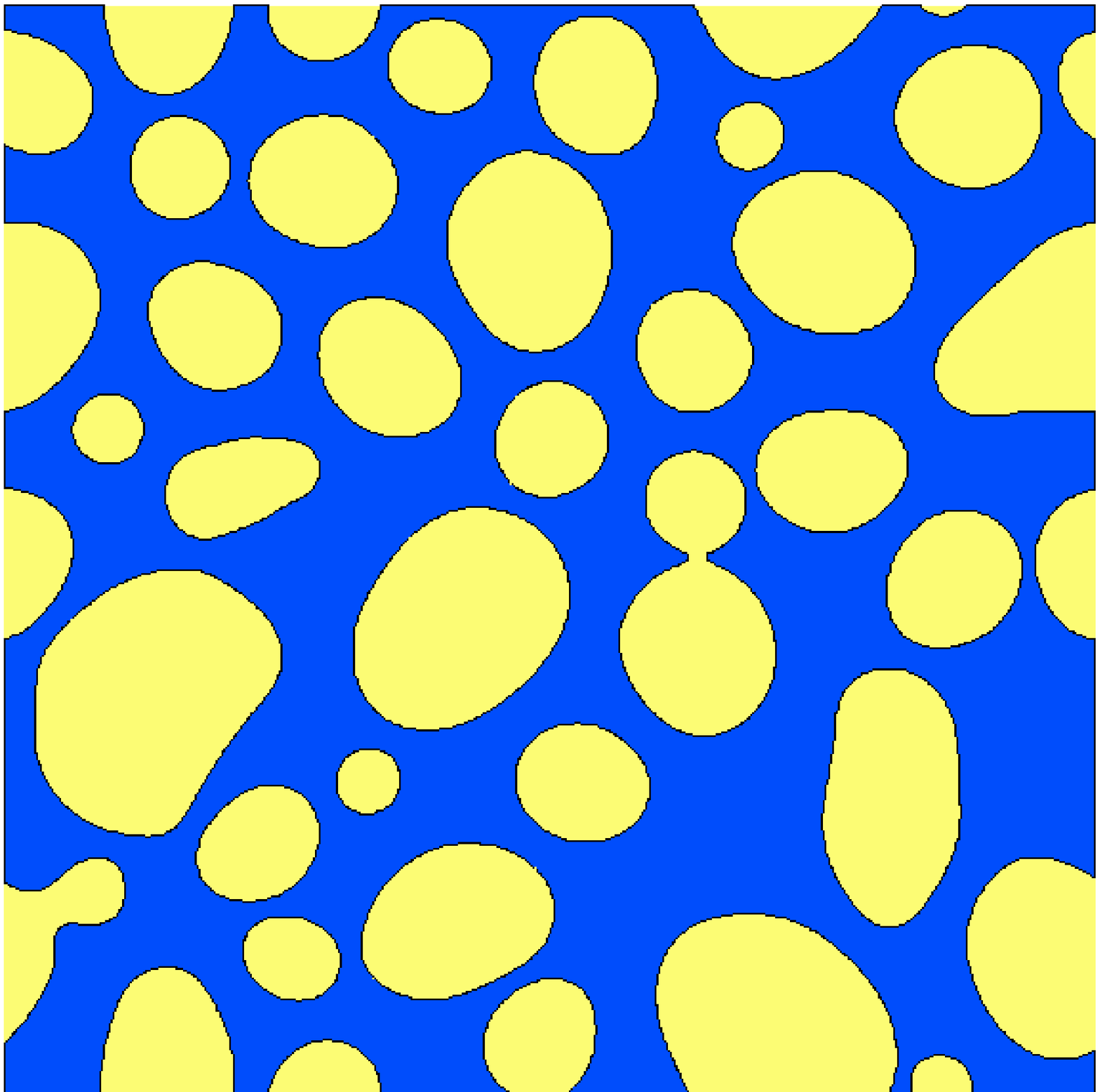,width=3cm}}
\end{minipage}
\hspace{0.3cm}
\begin{minipage}{3cm}
\centerline{\psfig{figure=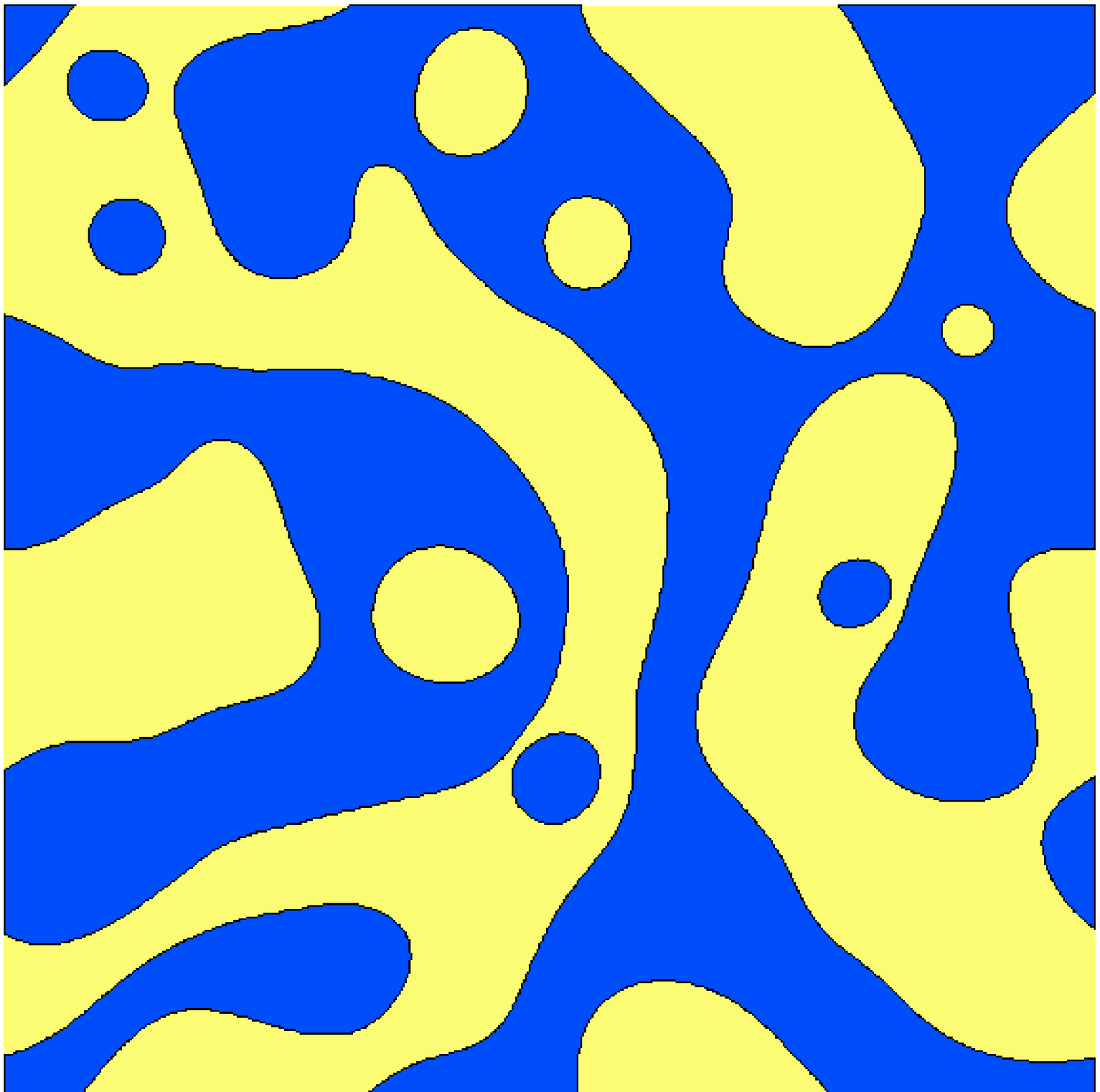,width=3cm}}
\end{minipage}
\begin{center}
(a) \hspace{2.5cm} (b) 
\end{center}
\begin{minipage}{6cm}
\centerline{\psfig{figure=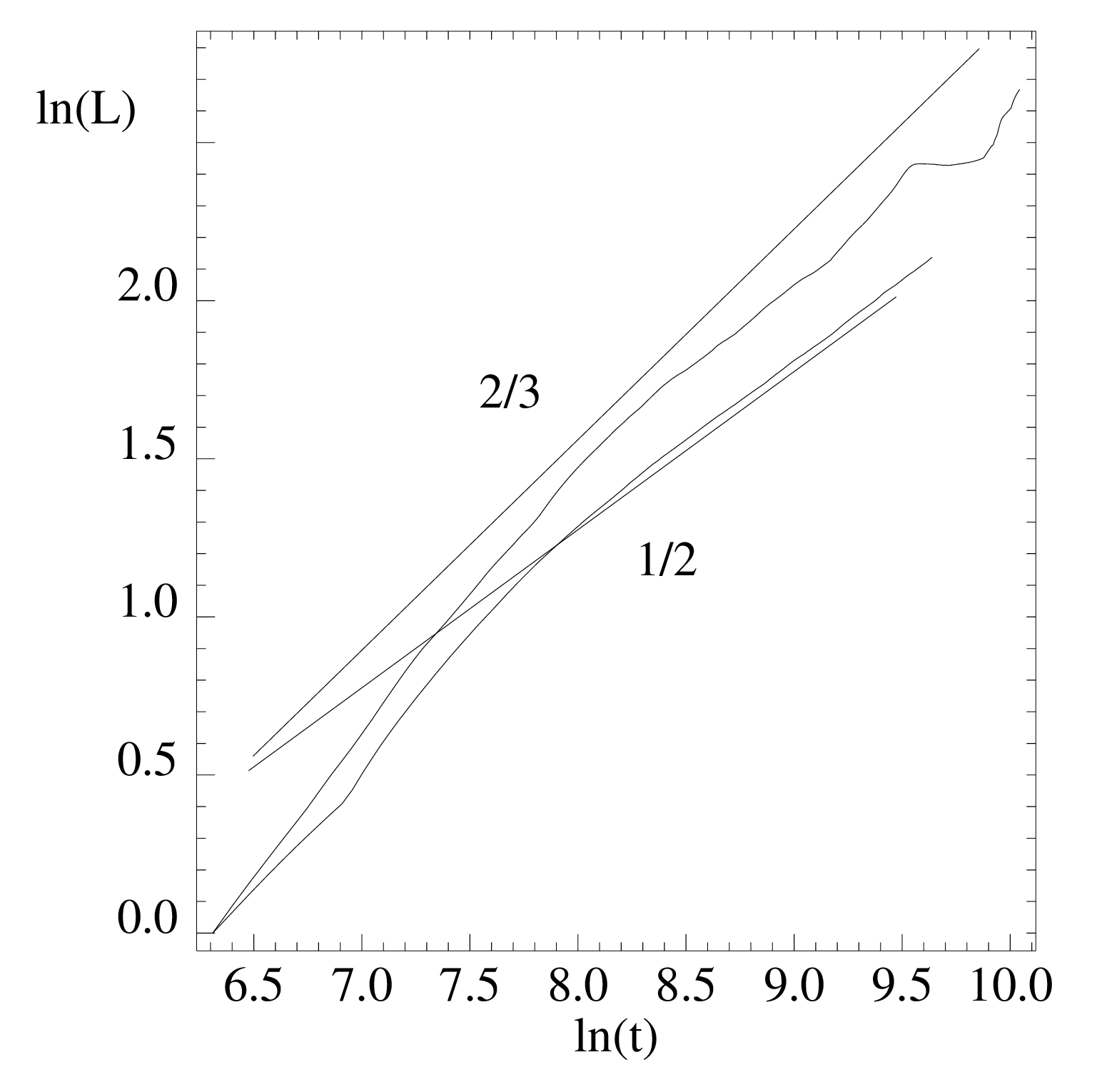,width=6cm}}
\end{minipage}
\begin{center}
(c)
\end{center}
\end{minipage}
\end{center}
\caption{A comparison of the phase-ordering of two identical symmetric
binary mixtures after different early time spinodal decomposition. In
(a) a phase-ordering is seen from an initial morphology generated by
viscoelastic phase separation in which the light component is
dispersed (the originally low-viscous component). In (b) we see the
usual symmetric phase separation. Fig.(c) shows the length scales
$L(t)$ for (a) $L\sim t^{1/2}$ and (b) $L\sim t^{2/3}$.}
\label{fig:comp}
\end{figure}

It is usual to assume that the phase-ordering process is independent
of the details of the early time spinodal decomposition, although this
assertion has, to our knowledge, only been tested with respect to
different random initial conditions. If this assumption were true we
would expect the domain growth to lead to a reconnection of the
dispersed domains that would lead to a morphology and a growth law
that are identical to a system which started with the same viscosities
for both components.

In Figure \ref{fig:comp} we see a comparison of the phase-ordering
from an initially dispersed phase (a) and a symmetric initial
condition (b). The morphologies for both systems are shown after 1000,
2000, 4000, and 8000 iterations. We see that the phase-ordering of the
morphology with a dispersed phase leads to an even more pronounced
dispersed phase where almost all domains are circular at late
times. We see that droplet coalescence occurs frequently with only few
domains vanishing due to the evaporation-condensation mechanism
underlying Oswald ripening. The droplet coalescence, however, is not
frequent enough to change the connectivity of the domains. Instead, we
observe that domains become more circular on average, suggesting that
even for very long times we do not expect a transition to a
bi-continuous morphology. In Figure \ref{fig:comp}(c) we see that the
growth law for the droplet phase is $L(t) \sim t^{1/2}$ and is smaller
than the $L(t)\sim t^{2/3}$ seen for the symmetric phase-ordering
shown in Figure \ref{fig:comp}(b), but also faster than the $L(t)\sim
t^{1/3}$ usually seen in the coarsening of off-critical droplet
morphologies of symmetric binary mixtures. (The data for the well
established scaling law of $L\sim t^{2/3}$ for the symmetric
phase-ordering in Figure \ref{fig:comp}(c) is more noisy since it
relies on a single simulation on a $256^2$ lattice where as the data for
the droplet morphology has been obtained by simulation on a $1024^2$
lattice.) We suggest that the high density of droplets
enhances the droplet coalescence an leads to a faster growth of $L\sim
t^{1/2}$. We measure $L(t)$ as the inverse of the length of the
interface. We should also point out that a length scale derived as
$L\sim 1/\sqrt{N}$, where $N$ is the number of domains, scales as
$L\sim t^{1/2}$ as well, suggesting that it may be a true scaling
state\cite{prl}.

The existence of this second scaling state, distinct from the
bi-continuous scaling state, is important to understanding the
late-time regime of viscoelastic phase separation because, even in the
absence of viscoelastic effects, we observe a droplet-morphology
evolving from the initial morphology created by viscoelastic phase
separation. This result is also important for practical reasons in
processes where late-time morphologies need to be controlled. It is
well known that mixing of high-viscosity and low-viscosity components
by means of mechanical agitation leads to morphologies where the
high-viscosity phase is dispersed in droplets\cite{Utracki} for volume
fractions of the low-viscosity component of much less than 50\%. This
effect is enhanced if the high-viscous phase is
viscoelastic\cite{Vanoene}. This leads us to consider what the
late-time morphology of a phase-ordering system with an early
morphology created by mechanical mixing would be.

\begin{figure}
\begin{center}
\begin{minipage}{7cm}
\begin{minipage}{3cm}
\centerline{\psfig{figure=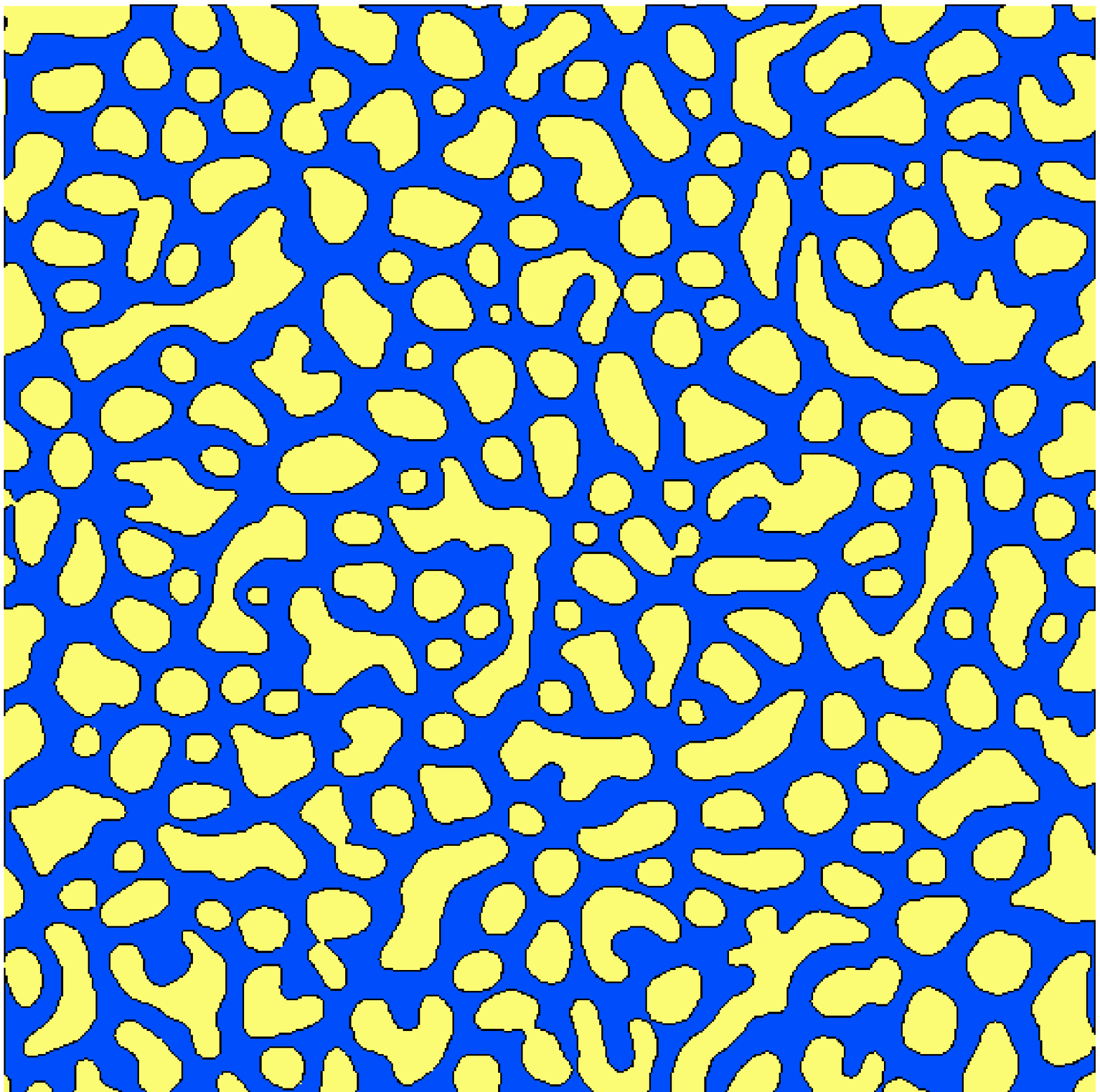,width=3cm}}
\end{minipage}
\hspace{0.3cm}
\begin{minipage}{3cm}
\centerline{\psfig{figure=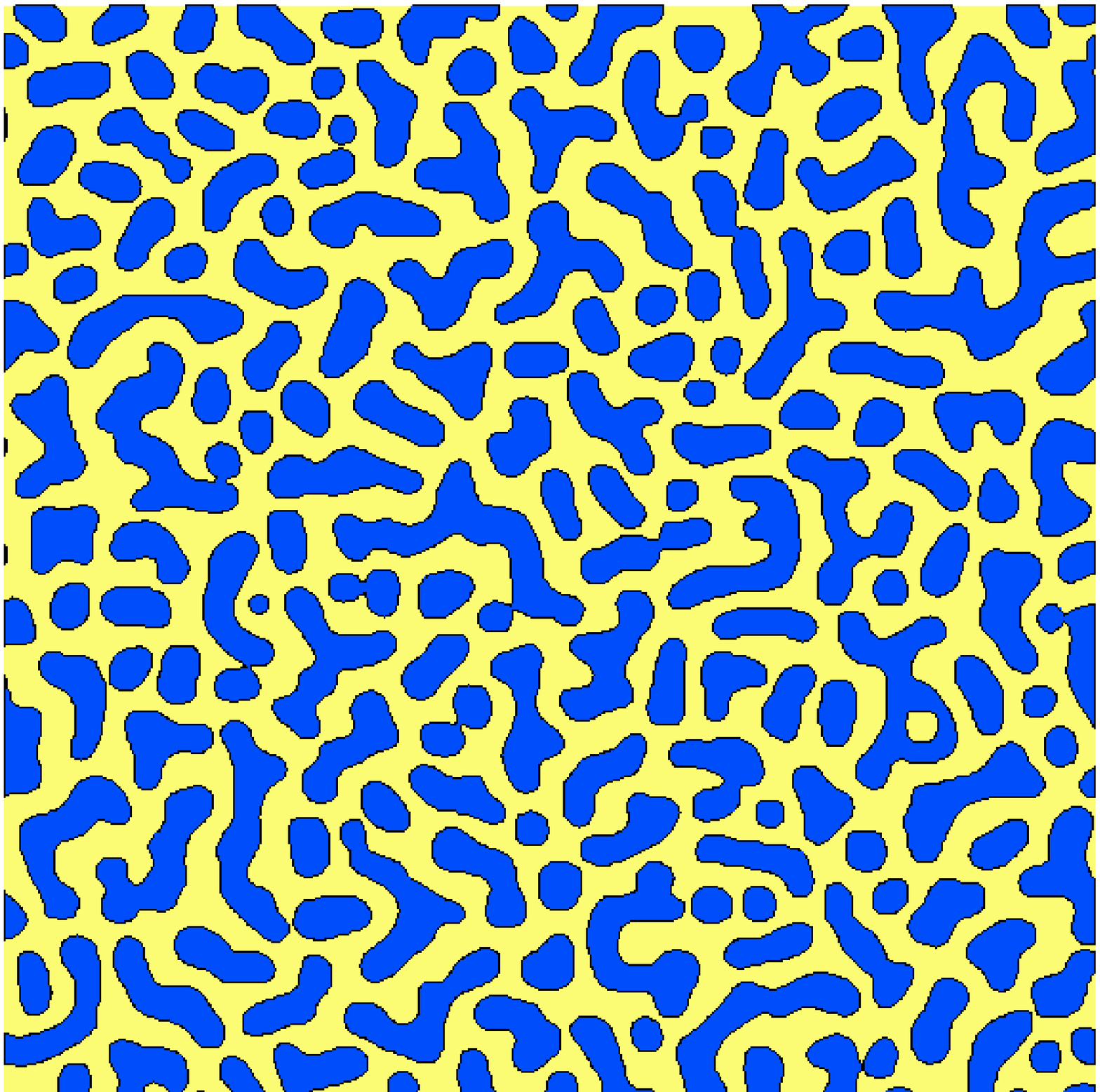,width=3cm}}
\end{minipage}\\
\begin{minipage}{3cm}
\centerline{\psfig{figure=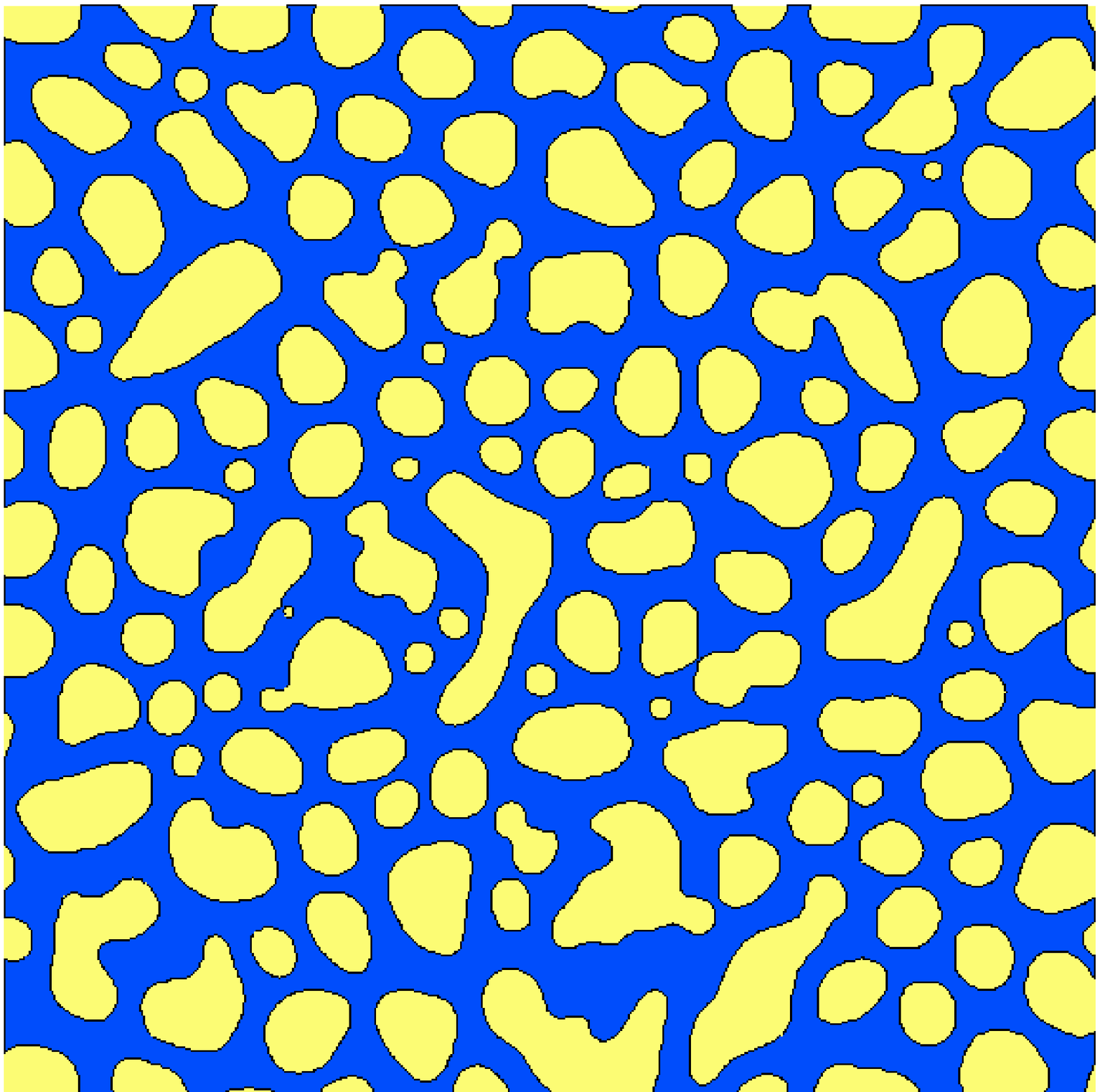,width=3cm}}
\end{minipage}
\hspace{0.3cm}
\begin{minipage}{3cm}
\centerline{\psfig{figure=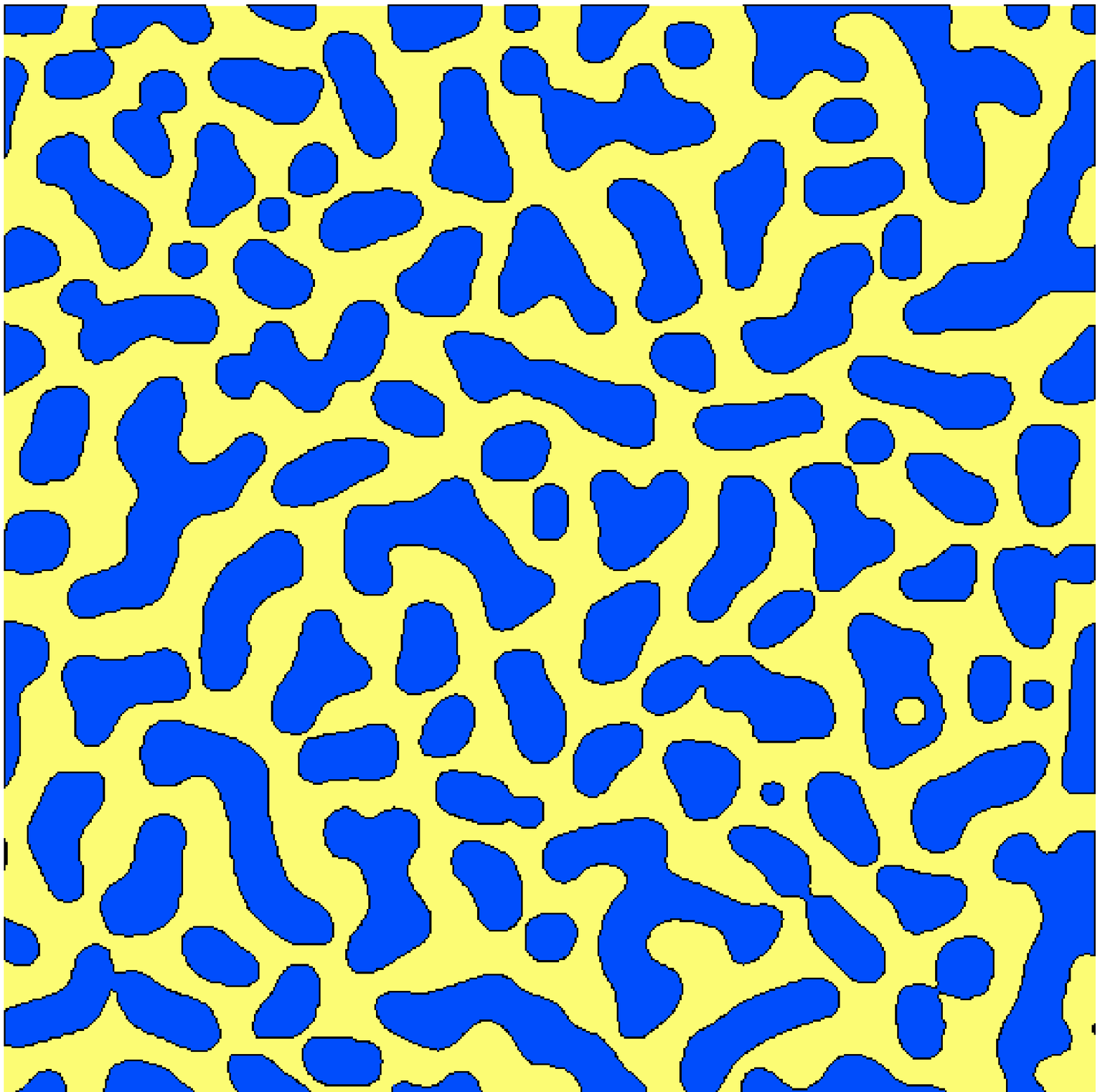,width=3cm}}
\end{minipage}\\
\begin{minipage}{3cm}
\centerline{\psfig{figure=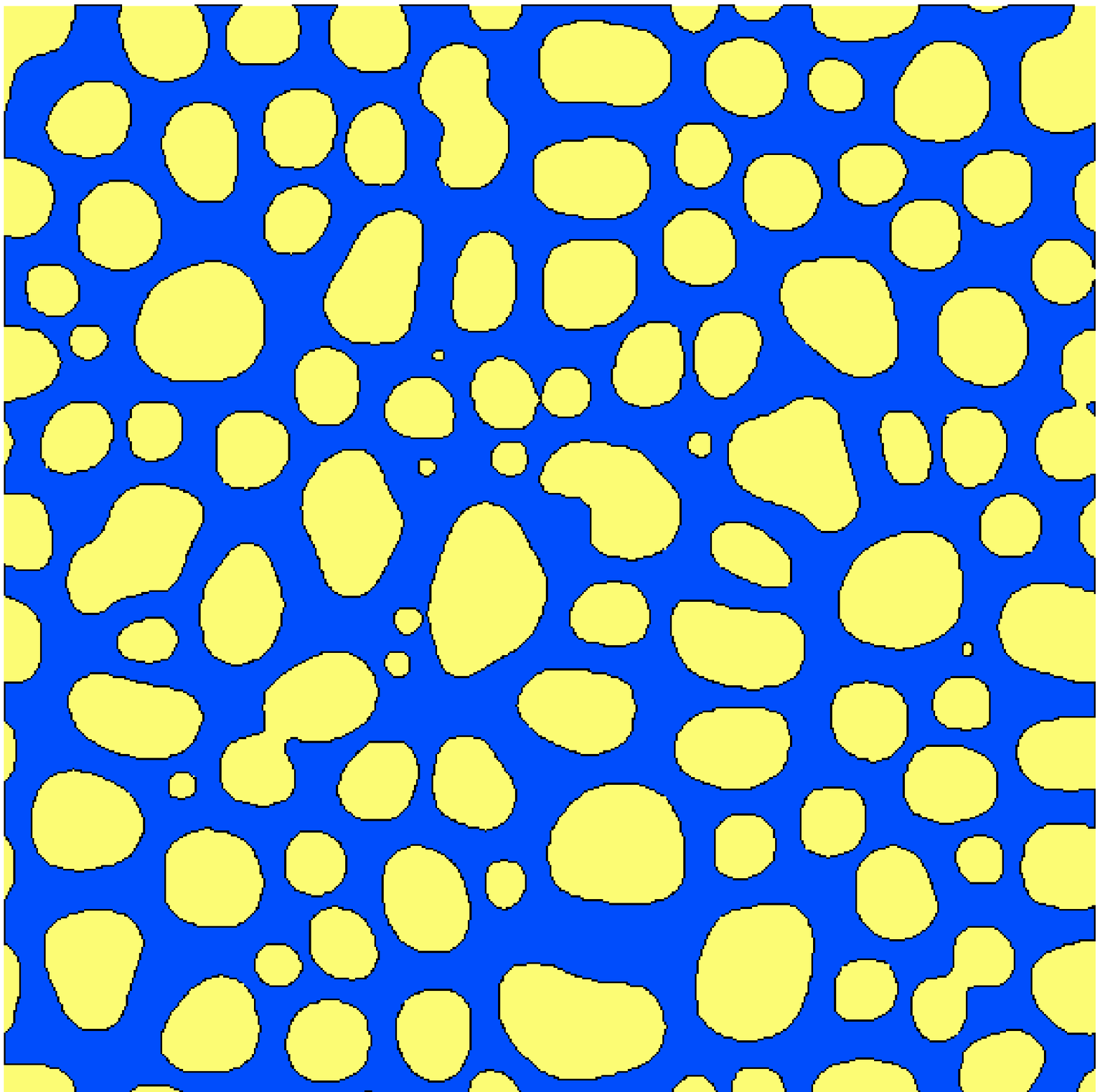,width=3cm}}
\end{minipage}
\hspace{0.3cm}
\begin{minipage}{3cm}
\centerline{\psfig{figure=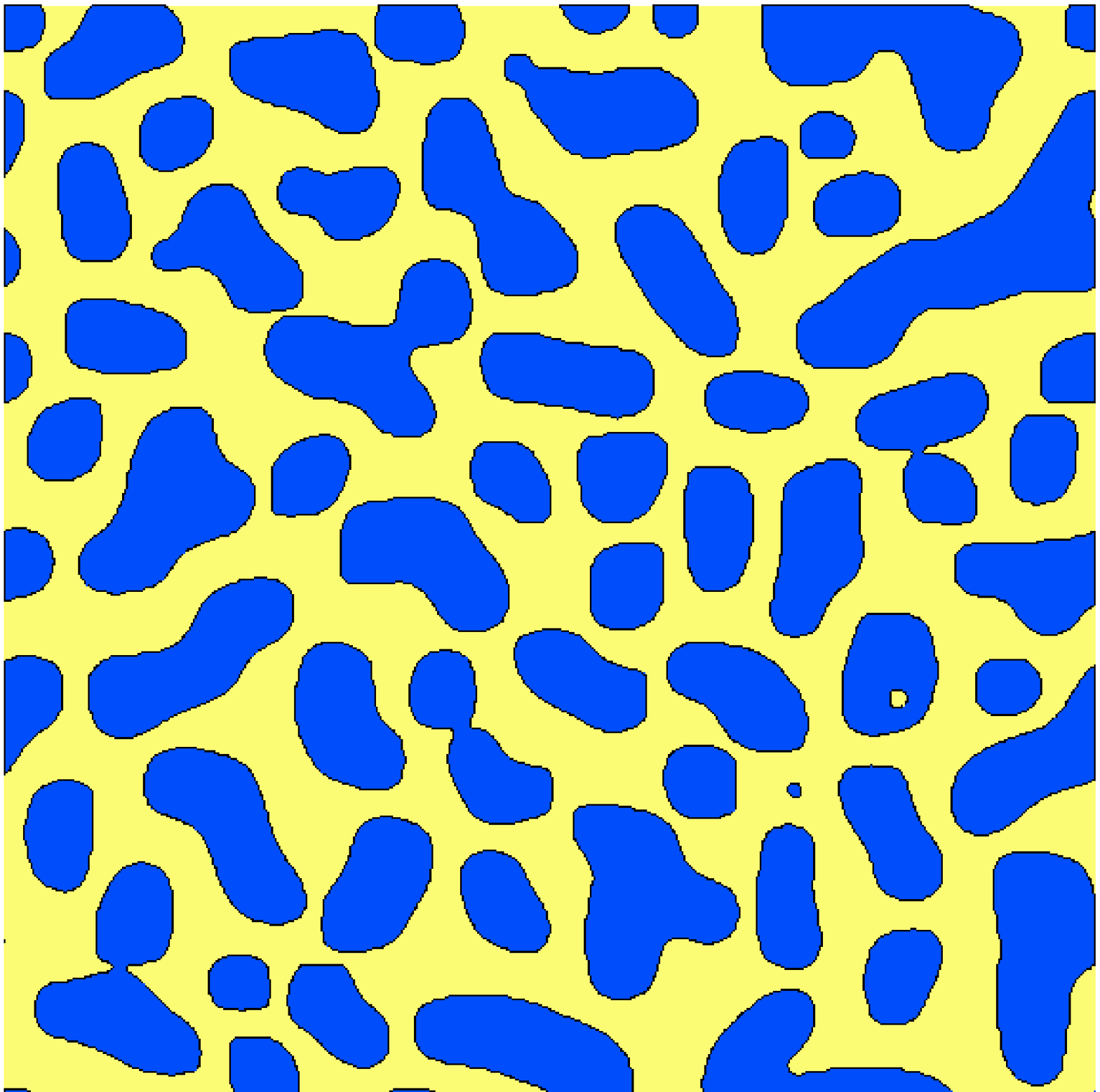,width=3cm}}
\end{minipage}\\
\begin{minipage}{3cm}
\centerline{\psfig{figure=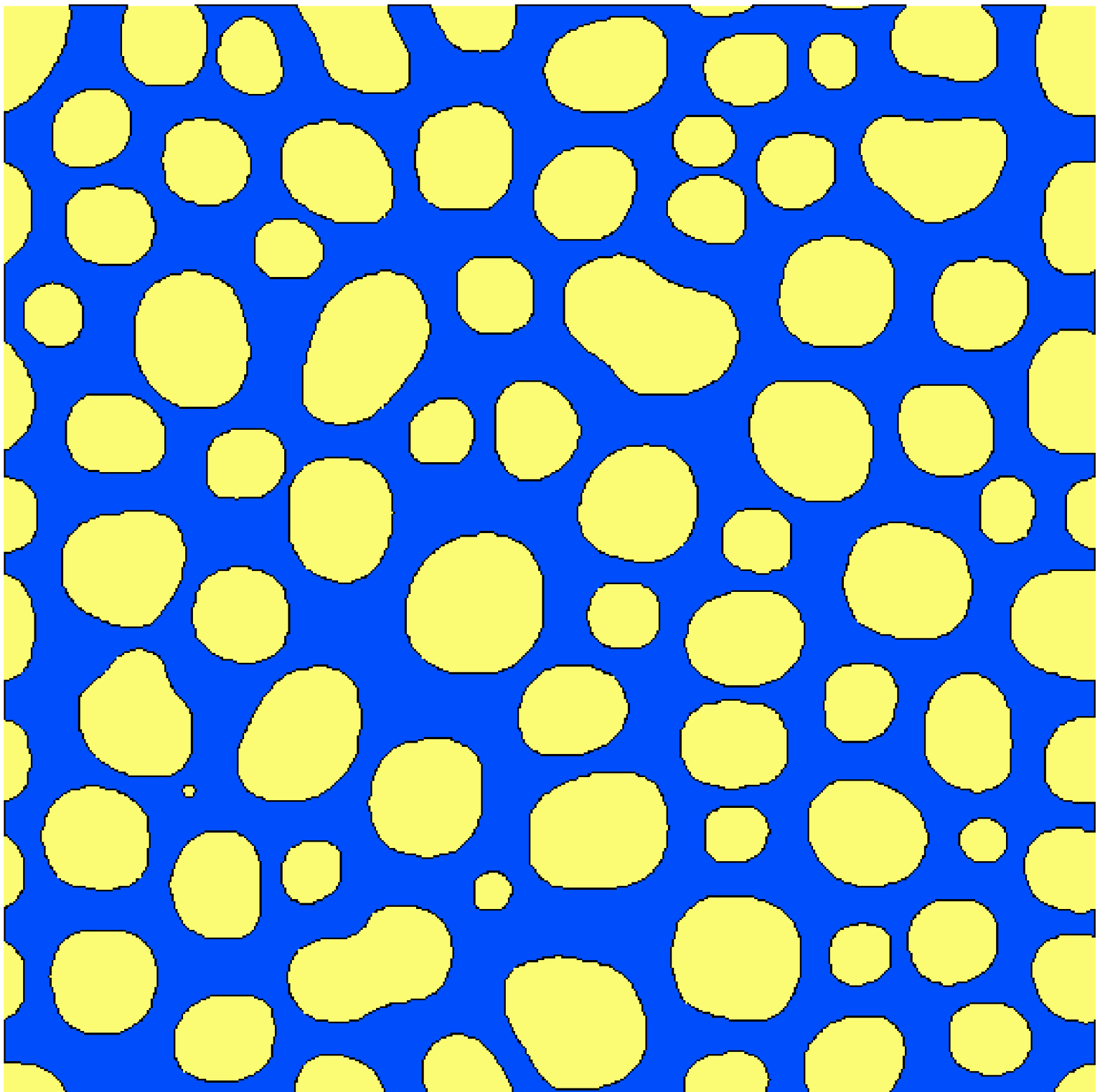,width=3cm}}
\end{minipage}
\hspace{0.3cm}
\begin{minipage}{3cm}
\centerline{\psfig{figure=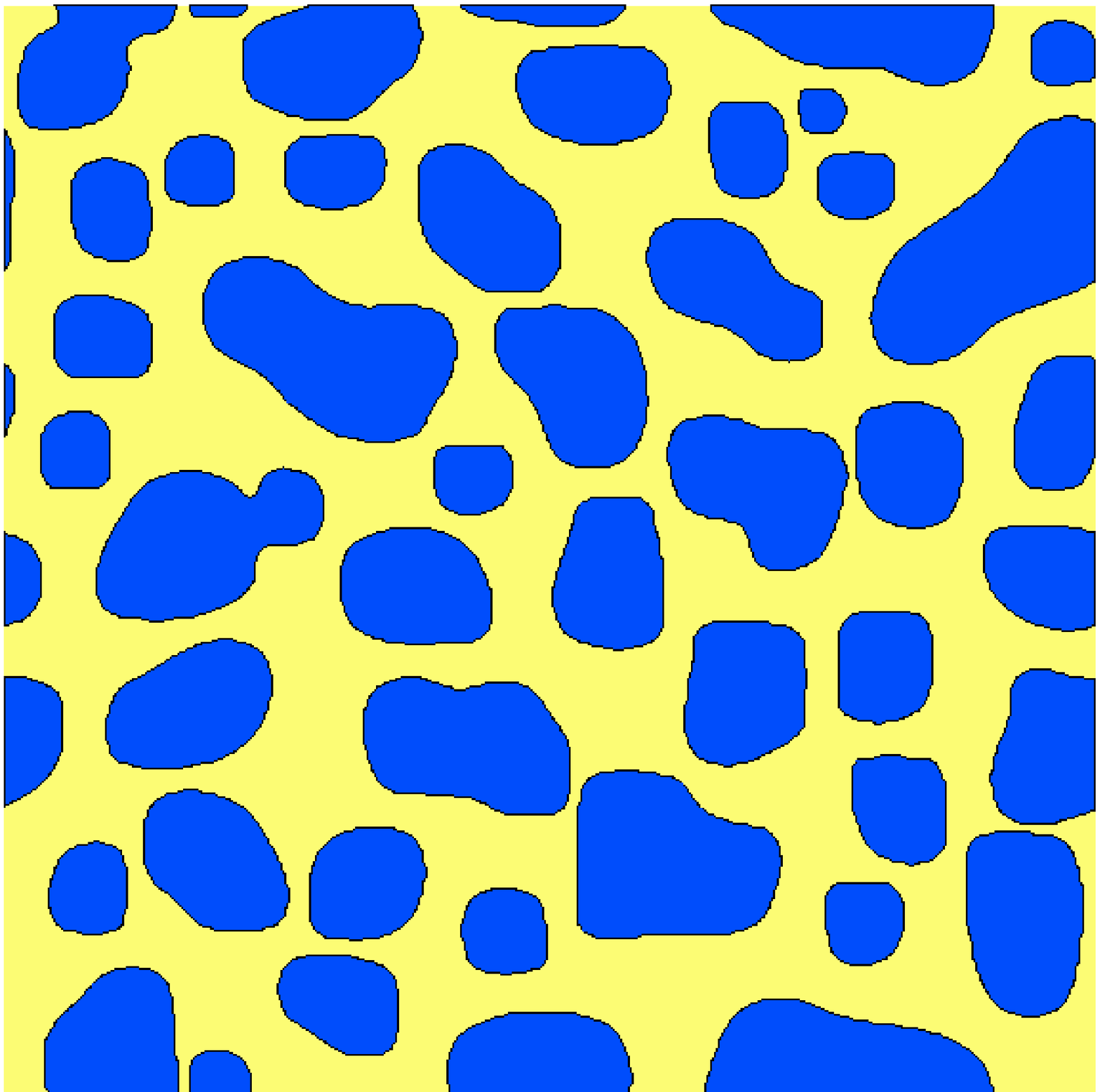,width=3cm}}
\end{minipage}
\begin{center}
(a) \hspace{2.5cm} (b) 
\end{center}
\begin{minipage}{6cm}
\centerline{\psfig{figure=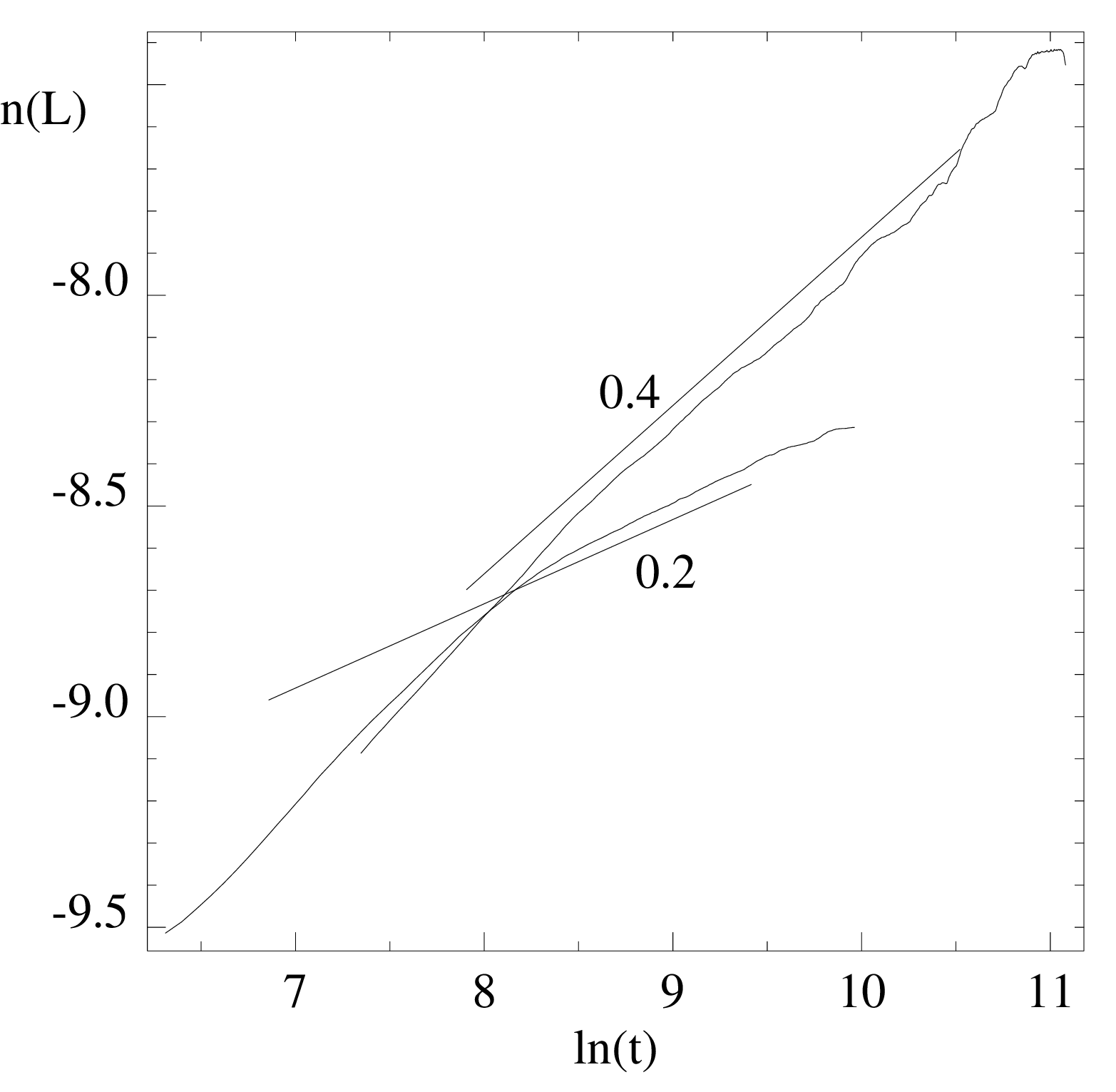,width=6cm}}
\end{minipage}
\begin{center}
(c)
\end{center}
\end{minipage}
\end{center}
\caption{The phase-ordering of a binary mixture of a viscoelastic
phase (dark color) and a low viscosity Newtonian phase (light color)
is shown in (a) and (b). (a) corresponds to a usual viscoelastic phase
separation of a 50\%-50\% mixture and (b) shows the phase-ordering
after the viscoelastic and low-viscosity domains have been
interchanged. Morphologies are shown after 2000, 4000, 8000, and 16000
iterations. The scaling behavior of the two mixtures is shown in (c)
where for (a) $L\sim t^{0.2}$ and for (b) $L\sim 0.4$.}
\label{fig:vecomp}
\end{figure}

In order to answer this question we performed a simulation of
viscoelastic phase separation and after a droplet morphology had been
formed at 1000 time-steps, we inverted the properties of the two
components. We used this state as a model for the morphologies of
dispersed droplets of the viscoelastic phase in a matrix of the low
viscosity Newtonian phase that is typical of mechanically mixed
morphologies. We then continued the simulations and observed the
phase-ordering behavior of the new morphology. The results of these
simulations are shown in Figure \ref{fig:vecomp}. In Figure
\ref{fig:vecomp}(a) the morphology for a viscoelastic phase separation
of a 50\%-50\% mixture is shown. We should emphasize that the
phase-ordering (see eqn. (\ref{diffusion})) does not have a
composition-dependent diffusion constant and therefore no domain
shrinkage is observed in these simulations. Domain shrinkage can be a
very slow process in viscoelastic phase separation that prolongs the
spinodal decomposition process and makes it more difficult to
differentiate the early-stage decomposition and the late-time
phase-ordering processes. This does not reduce the validity of our
results, however, since we are only interested in the late-time
behavior when the domain shrinkage is completed.

The morphologies shown in Figure \ref{fig:vecomp}(b) are of a
simulation where at 1000 iterations, after the dispersion was
achieved, the viscoelastic and the low viscosity components were
exchanged. We see that the morphology remains a droplet morphology,
albeit now of the viscoelastic phase. Comparing Figures
\ref{fig:vecomp} (a) and (b) we see that the most important factor in
determining the final morphology is the early stage dynamics, and that
phase-ordering scaling morphologies of both dispersed viscoelastic
and dispersed low viscosity domains exist. 

From Figure \ref{fig:vecomp}(c) we see that each scaling state has a
different growth law. The morphology of dispersed viscoelastic domains
grows as $L\sim t^{0.4}$ whereas the morphology of dispersed low
viscosity domains grows as $\sim t^{0.2}$ after near circular droplets
have been formed. An anomalously slow growth  in viscoelastic phase
separation has first been observed experimentally by
Tanaka\cite{tanaka} who found $L\sim t^{0.15}$ for a system of
high molecular-weight polystyren/ diethyl malonate (4.0 wt.\%).

These simulations also emphasize the difference between a morphology
after spinodal decomposition and after mechanical mixing. The
morphology after spinodal decomposition is a dispersed low viscosity
phase, whereas the state after mechanical mixing has a dispersed
viscoelastic phase. Subsequent phase-ordering does not change the
connectivity of these states, in agreement with the conventional
wisdom that there is a profound difference in states produced by
spinodal decomposition and mechanical mixing.

\section{Conclusions}
In this letter we have shown that more than one scaling state exists
for late-time spinodal decomposition and that the early time
behavior of a phase-separating binary mixture is very important in
selecting one of these scaling states. We have also explained the
difference between viscoelastic phase-ordering states after spinodal
decomposition and mechanical mixing. Our results show that the volume fraction
and the physical properties of a mixture do not select a morphology by
themselves, but that the morphology of the initial state is of
paramount importance. This is why viscoelastic phase separation can
lead to unusual late-time scaling states even when viscoelasticity is
no longer important at large length scales.

\section*{Acknowledgments}
The authors acknowledge the financial support of DuPont Chemical
Company. A.W. would like to thank Craig Carter for the generous
permission to use his Origin2000 computer and Heidi Burch for editing
the manuscript.

\def\jour#1#2#3#4{{#1} {\bf #2}, #3 (#4)}.
\def\tbp#1{{\em #1}, to be published}.
\def\inpr#1{{\em #1}, in preparation}.
\def\tit#1#2#3#4#5{{#1} {\bf #2}, #3 (#4)}

\def\ap{Adv. Phys.}
\def\arf{Ann. Rev. Fluid Mech.}
\def\epl{Euro. Phys. Lett.}
\def\ijmp{Int. J. Mod. Phys. C}
\def\jcp{J. Chem. Phys.}
\def\jpc{J. Phys. C}
\def\jpcs{J. Phys. Chem. Solids}
\def\jpco{J. Phys. Cond. Mat}
\def\jsp{J. Stat. Phys.}
\def\jf{J. Fluids}
\def\jfm{J. Fluid Mech.}
\def\jnnfm{J. Non-Newtonian Fluid Mech.}
\def\pfa{Phys. Fluids A}
\def\prl{Phys. Rev. Lett.}
\def\pr{Phys. Rev.}
\def\pra{Phys. Rev. A}
\def\prb{Phys. Rev. B}
\def\pre{Phys. Rev. E}
\def\pa{Physica A}
\def\pla{Phys. Lett. A}
\def\ps{Physica Scripta}
\def\roy{Proc. Roy. Soc.}
\def\rmp{Rev. Mod. Phys.}
\def\zpb{Z. Phys. B}

\end{document}